\documentclass[aps,prl,reprint,groupedaddress]{revtex4-1}
\usepackage{graphicx, color, soul}

\begin{document}
\title{Photonic Realization of a Transition Between Topological Phases Residing in the Weak and Strong Floquet Driving Regimes}

\author{Jonathan Guglielmon$^{1}$, Sheng Huang$^{2}$, Kevin P. Chen$^{2}$, and Mikael C. Rechtsman$^{1}$}
\affiliation{$^{1}$Department of Physics, The Pennsylvania State University, University Park, PA 16802, USA\\\mbox{$^{2}$Department of Electrical and Computer Engineering, University of Pittsburgh, Pittsburgh, Pennsylvania 15261, USA}}

\date{\today}

\begin{abstract}
A photonic Floquet topological insulator has previously been experimentally realized in an array of evanescently-coupled helical waveguides.   In the topological regime probed by that experiment, the chirality of the single topological edge mode was the same as the chirality of the helices. Here we demonstrate the rich structure present in Floquet systems by fixing the helix chirality while moving into the strong driving regime to observe a topological transition in which the edge mode reverses its propagation direction, yielding the counterintuitive result that an increase of the driving amplitude can have the same effect as reversing the helix chirality. We experimentally observe this transition while overcoming the bending loss typically associated with the strong driving regime, despite the fact that the helix radius is on the same scale as the size of the entire array. The two topological phases observed in our experiment can be understood as originating from a Floquet realization of the Haldane model.
\end{abstract}

\maketitle
In recent years, the field of condensed matter physics has been profoundly impacted by the discovery of topological insulators, a state of matter in which the global, topological structure of the system's eigenstates results in surprisingly robust properties. Though originally discovered in the condensed matter context, many of the underlying topological ideas have since been realized in a variety of other settings including photonic \cite{haldane2008possible,wang2009observation,rechtsman2013photonic,hafezi2013imaging,lu2014topological,cheng2016robust}, ultra-cold atomic \cite{jotzu2014experimental,aidelsburger2015measuring,goldman2016topological}, and mechanical systems \cite{prodan2009topological,zhang2010topological,kane2014topological,nash2015topological,huber2016topological}. In addition to potentially enabling unique device functionalities within these fields, the advent of topological physics in these settings provides a platform for the experimental realization of topological phenomena in a context where it is possible to directly engineer the microscopic details of the system, including the underlying lattice, the interactions, and the structure of any applied gauge fields. 

A particularly interesting example in photonics is provided by paraxial waveguide arrays \cite{szameit2010discrete}, where the physics of paraxial light diffracting through a collection of evanescently-coupled waveguides is identical to the physics of a non-interacting electron confined to two dimensions evolving according to the Schr\"{o}dinger equation. While for electrons the Hamiltonian generates evolution in time, for photons it generates evolution along the paraxial spatial direction (i.e., the propagation axis of the waveguides). This map from temporal to spatial degrees of freedom can be exploited to explore rich Floquet phenomena associated with intricate time-dependent Hamiltonians.  This provides a particularly fruitful avenue for obtaining topological systems in photonics, since it is known that Floquet systems can exhibit topologically non-trivial phases \cite{oka2009photovolatic,kitagawa2010topological,lindner2011floquet,gu2011floquet}.  Furthermore, this gives rise to novel effects that can be explored in the context of Floquet topological physics related to the fact that photons are bosons, and that photonic systems are by nature strongly out of equilibrium.

A photonic realization of such a Floquet topological insulator was given in \cite{rechtsman2013photonic} where a honeycomb array of helical waveguides was fabricated such that the waveguide helicity generates an effective gauge field that drives the system to a topologically non-trivial phase. In the regime probed by that experiment, it is the chirality of the helices that determines the chirality of the topological edge mode. In particular, if the waveguides spiral clockwise (counterclockwise), then the edge mode will likewise circulate clockwise (counterclockwise) around the boundary of the sample.

However, being a Floquet system, there are additional degrees of freedom associated with the driving field that can potentially lead to more elaborate topological behavior. In particular, even for a fixed helix chirality, one can tune the amplitude and frequency of the effective gauge field. These additional parameters provide a continuous two-dimensional parameter space that has been shown, in the condensed matter context, to result in a surprisingly rich phase diagram \cite{mikami2016brillouin}. 

In this letter, we probe the topological landscape generated by the amplitude degree of freedom. In particular, we experimentally demonstrate that by moving into the strong driving regime -- that is, the regime of large helix radius -- we can enter a phase in which the Floquet topological winding number \cite{rudner2013anomalous} changes sign and the associated topological edge mode reverses its propagation direction so that its chirality is opposite to the chirality of the helices (we use the winding number here instead of the Chern number since it is the appropriate bulk invariant for Floquet systems \cite{rudner2013anomalous}). A complication encountered in the strong driving regime is the problem of waveguide bending loss (equivalent to the problem of overheating in condensed matter), an effect that inhibited observation of a phase transition in prior experimental studies \cite{rechtsman2013photonic}.  We find that this problem can be circumvented by working in the highly irregular parameter regime in which the helix radius is roughly the size of the entire waveguide array. While a photonic topological transition has been demonstrated previously \cite{leykam2016anamolous,noh2017experimental} (from topological to trivial), the model we study here is clearly distinct in the sense that it can be mapped in an appropriate limit to the Haldane model (i.e., the quantum anomalous Hall effect \cite{haldane1988model,chang167experimental}), highlighting its similarities to the Haldane model while showing that the Floquet system is in fact much richer.  Furthermore, it has direct predictive implications for the mathematically equivalent condensed matter system of graphene irradiated by strong, circularly polarized light \cite{oka2009photovolatic,gu2011floquet}.  In particular, we show -- by analogy with photonics -- that the strength of the gauge field (i.e., intensity of the irradiating light) can be used to tune between topological phases.  Furthermore, like the electronic system, the photonic system is also constrained by `heating' (i.e., bending loss), and we provide a prescription for overcoming its limitations.

\begin{figure}
\includegraphics[width=\linewidth]{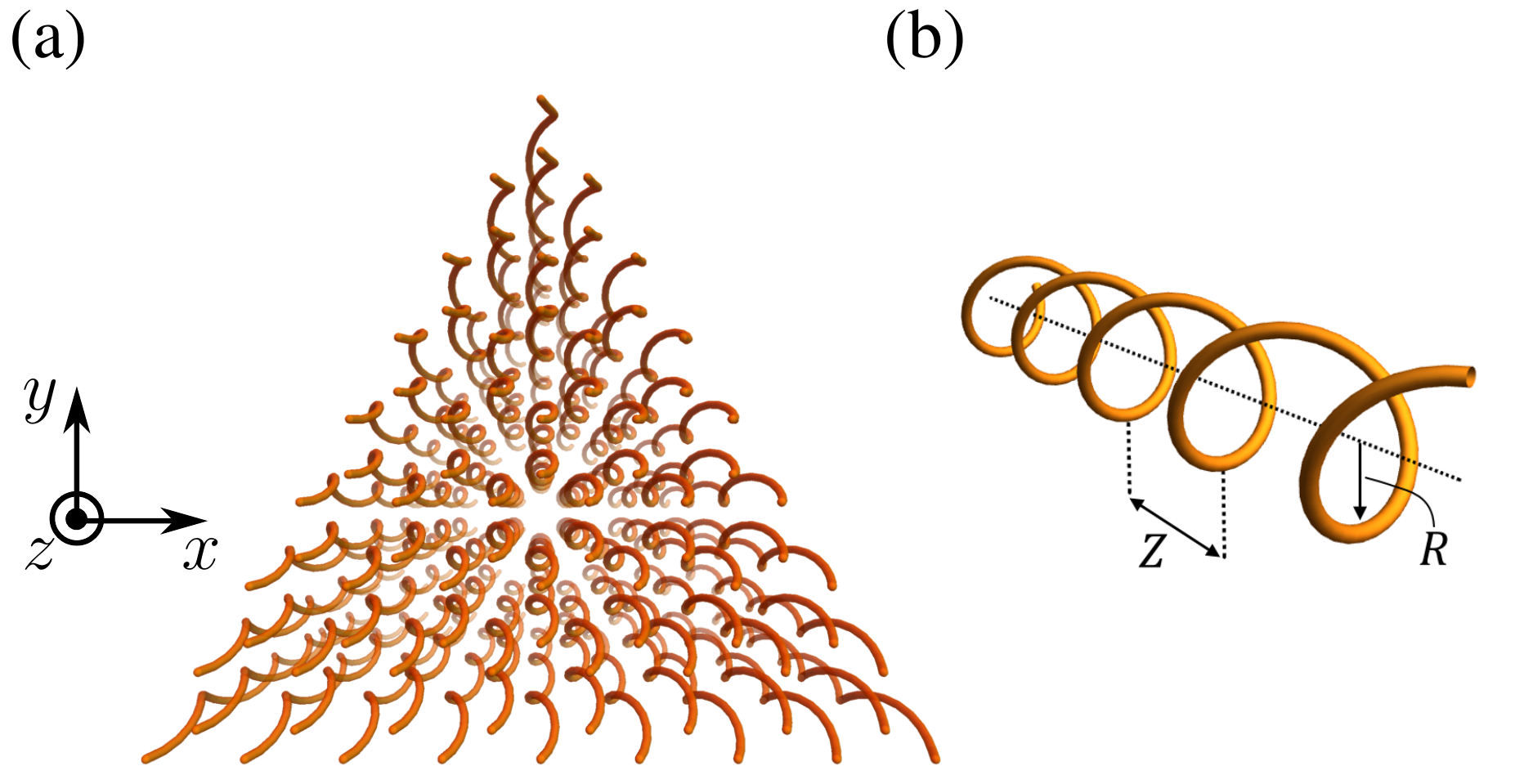}
\caption{\label{fig_lattice} (a) Illustration of our honeycomb array of helical waveguides. Light is injected at the front of the structure and evolves in the transverse plane as it propagates into the page along the $z$-direction. (b) An isolated waveguide highlighting the parameters $(R,Z)$. By varying these parameters, we can place the system in a variety of topological phases.}
\end{figure}

Figure \ref{fig_lattice} illustrates our photonic structure, which consists of a honeycomb array of helical waveguides aligned along the $z$-direction. The helices are characterized by their radius $R$ and their spatial period $Z$. We also define the helix frequency $\Omega = 2 \pi/Z$. In the paraxial approximation, the electric field $\mathbf{E}(x,y,z) = \psi(x,y,z)\exp(i k_0 z - i\omega t)\hat{\mathbf{E}}_0$ is governed by an equation resembling the Schr\"{o}dinger equation, in which the paraxial direction, $z$, takes the place of time and the variation, $\delta n$, of the refractive index $n = n_0 + \delta n$ plays the role of a potential. Here $\omega=2\pi c/\lambda$ is the operating frequency, $\lambda$ is the wavelength, and $k_0 = 2 \pi n_0 /\lambda$ is the background wavenumber. The waveguides used in our experiment have been engineered to exhibit a single bound mode each for wavelengths in the vicinity of $1.55 \mu m$. We choose our lattice constant so that the resulting paraxial Schr\"{o}dinger equation can be modeled using tight-binding theory where light hops between the bound modes of adjacent waveguides via evanescent coupling.  The effect of the helices is to introduce a $z$-dependent U(1) gauge field \cite{rechtsman2013photonic}
\begin{equation}
\mathbf{A}(z) = k_0 R \Omega (\sin \Omega z, -\cos \Omega z, 0)
\end{equation}
 that modifies the hopping amplitudes with a Peierls phase yielding a tight-binding Schr\"{o}dinger equation
\begin{equation}\label{eqn_tight_binding}
i \partial_z \psi_n(z) = \sum_{\langle m\rangle} c e^{i \mathbf{A}(z) \cdot \mathbf{r}_{m n} } \psi_m(z)
\end{equation}
where $\psi_n$ is the amplitude of the electric field in the $n^{\text{th}}$ waveguide, $\mathbf{r}_{mn}$ is the displacement between sites $m,n$, $c$ is the hopping constant, and the sum over $m$ is taken over nearest neighbors. We will denote the eigenvalues of the Hamiltonian associated with Equation \ref{eqn_tight_binding} by $\beta$.

\begin{figure}
\includegraphics[width=\linewidth]{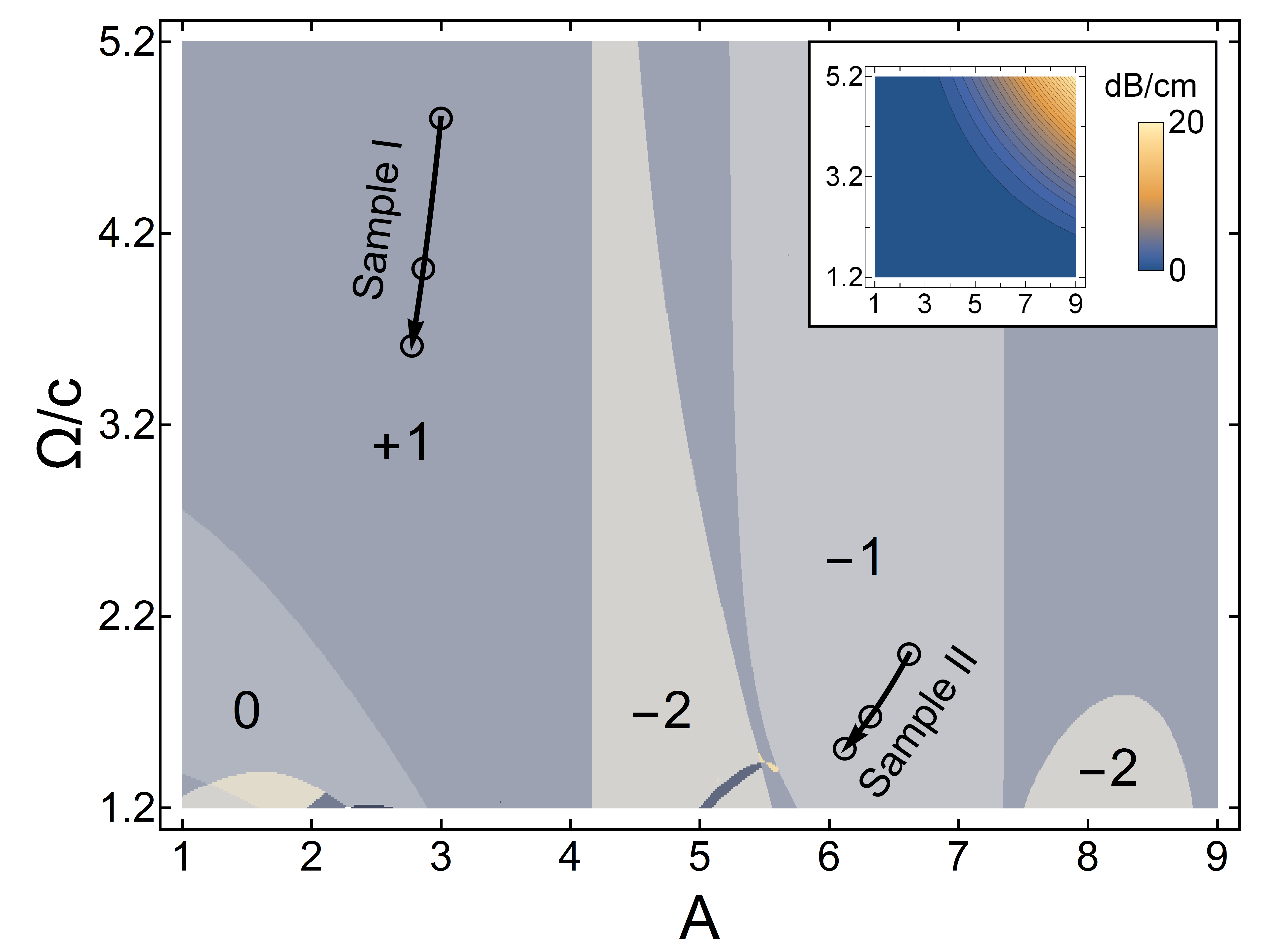}
\caption{\label{fig_phasediag} Floquet topological phase diagram showing the winding number associated with the $\beta=0$ gap as a function of the dimensionless parameters $\Omega/c$ and $A$. Inset shows a theoretical estimate of the bending loss plotted on axes identical to those of the phase diagram. The regions of the phase diagram probed in our experiment are labeled `Sample I' and `Sample II' alongside an arrow that shows the path taken by a wavelength sweep from $1480$-$1600$nm. The three points highlighted along the path correspond to the wavelengths of Figure \ref{fig_output}. }
\end{figure}

In the absence of the gauge field, the band structure for this system reduces to that of graphene and possesses two distinct Dirac cones at the corners of the Brillouin zone.  The introduction of $\mathbf{A}(z)$ breaks $z$-reversal symmetry and is thus capable of driving the system to topologically non-trivial phases. Since this is a Floquet system, the appropriate topological invariant is the winding number introduced in \cite{rudner2013anomalous}, which we compute for the gap centered on $\beta=0$. Here the winding number is fully determined by the two dimensionless parameters $\Omega/c$ and $A = ak_0R\Omega$, which correspond to the frequency and amplitude of the gauge field. The resulting phase diagram is shown in Figure \ref{fig_phasediag}, which was computed using the truncated Floquet scheme given in \cite{rudner2013anomalous} combined with the algorithm of \cite{fukui2005chern}. Note that due to the close relation between our photonic system and the Schr\"{o}dinger equation, this is the same Floquet topological phase diagram that appears when studying graphene irradiated by circularly-polarized light \cite{mikami2016brillouin}.

For this system, the gauge field driving amplitude is a function of both the helix radius and helix frequency: $A =ak_0 R\Omega$. Thus, for a fixed helix frequency, an increase in the amplitude $A$ will result in a decrease in the curvature radius $R_c = 1/(R \Omega^2)$
of the waveguides (note the distinction between the curvature radius $R_c$ and the helix radius $R$). In general, waveguide bending loss increases as $R_c$ is decreased \cite{marcuse1976curvature} and it was precisely these losses that prohibited the observation of a phase transition in \cite{rechtsman2013photonic}. A key result of this current work is that, by increasing the gauge field amplitude while simultaneously reducing its frequency, the losses can be reduced to a degree that allows us to observe a new topological phase. 

To determine which regions of the phase diagram are excluded from experimental observation by high bending loss, we show in the inset of Figure \ref{fig_phasediag} the bending loss computed over the same parameter space used in the plotting the phase diagram. We note that in mapping the loss over this parameter space, we have assumed a lattice constant of $a=22\sqrt{3} \mu m$. These losses represent a theoretical estimate computed using the result of \cite{marcuse1976curvature} for light of wavelength $1.55\mu m$. By working in the lower region of the phase diagram, we can reduce the losses to a degree that enables observation of new topological phases.

To the best of our knowledge, the only region of this phase diagram that has been realized experimentally in a photonic system is the low amplitude $W=+1$ region \cite{rechtsman2013photonic}. In this paper, we are concerned with whether we can realize a new phase, residing in the strong driving regime, for which the relation between the edge mode chirality and the waveguide chirality is reversed compared to the $W=+1$ phase. From Figure \ref{fig_phasediag}, we see that such a transition can be achieved by increasing the effective gauge field amplitude to move into a region with either $W=-2$ or $W=-1$. In this paper, we will restrict our attention to the observation of the $W = \pm 1$ regions of the phase diagram and leave observation of the higher winding number phases to future experiments. 

We note that these $W=\pm1$ phases have a close relation to the two non-trivial phases of the Haldane model \cite{haldane1988model}. In particular, they persist at arbitrarily high frequencies where they can be understood by examining the inverse frequency expansion of the effective Floquet Hamiltonian \cite{bukov2015universal}, which reproduces the Hamiltonian of the Haldane model with an inversion symmetry breaking mass $M=0$ and a time-reversal symmetry breaking parameter  $\phi = \text{sgn}(f)\pi/2$, with \cite{mikami2016brillouin}
\begin{equation}
f(A) =  \sum_{m\ne0} \frac{J_m^2(A/\sqrt{3}) \sin(2m\pi/3)}{m}
\end{equation} 
where $J_m(x)$ are the Bessel functions of the first kind. As a result, the winding number evaluates to either $\pm1$ and is selected by the sign of $f(A)$, which is in turn controlled by the amplitude of $\mathbf{A}(z)$. 

\begin{figure}
\includegraphics[width=0.8\linewidth]{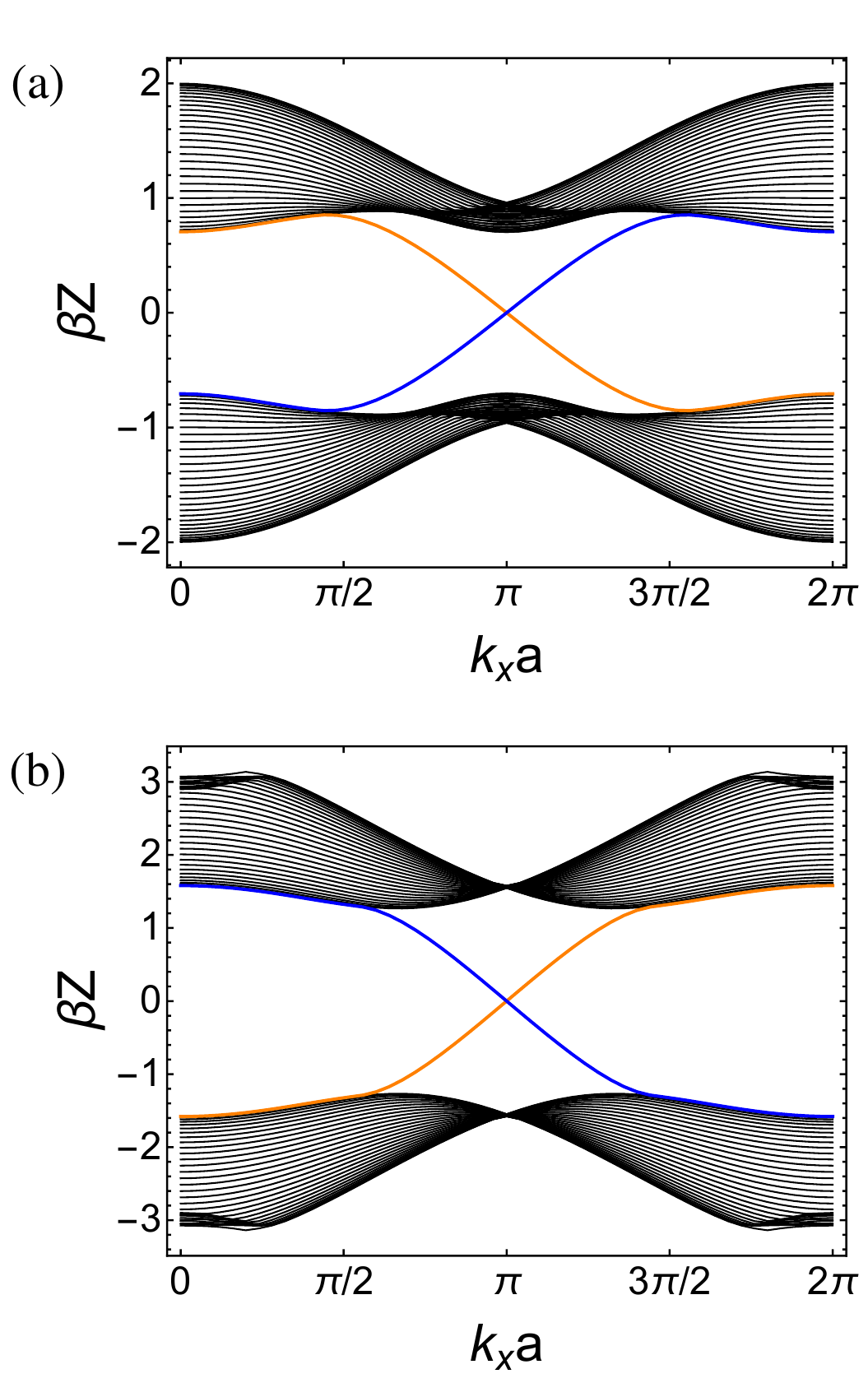}
\caption{\label{fig_bandstructure} Band structures for samples taken to be finite along the $y$-direction and periodic along the $x$-direction. The $W=+1$ phase is shown in (a) and the $W=-1$ phase in (b). These band structures are evaluated at the points in the phase diagram that correspond to the locations of the samples used in the experiment when operating at a wavelength of $1.55\mu m$. Edge modes highlighted in blue (orange) are localized on the bottom (top) of the sample. The interchange of blue/orange between the two band structures indicates the change in edge mode chirality expected from the sign change of the associated topological invariant.}
\end{figure}

\begin{figure*}	
\includegraphics[width=\linewidth]{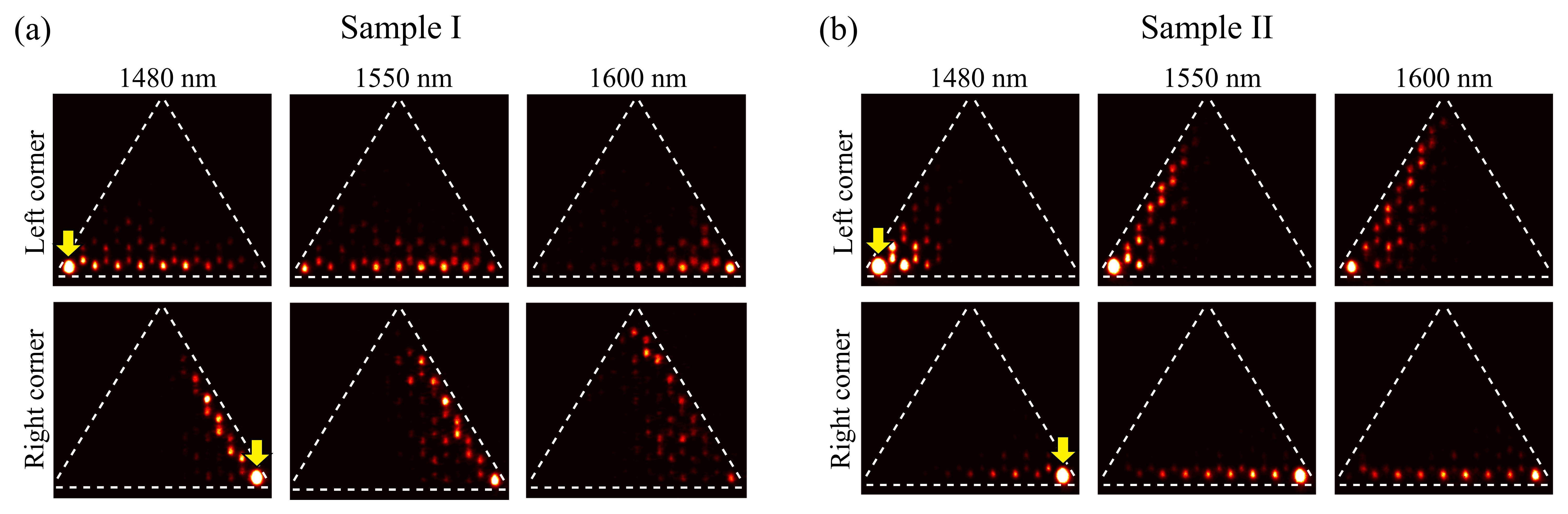}
\caption{\label{fig_output} Light arriving at the output facet after $14$cm of propagation. Panels (a) and (b) correspond, respectively, to samples that have been placed in the $W=+1$ and $W=-1$  regions of the phase diagram. Light is injected at the corner indicated by the arrow: the first (second) row corresponds to injection at the left (right) corner. Dashed white lines have been overlaid on the images to indicate the sample boundaries. By sweeping the wavelength, we effectively observe the light at different propagation distances along the sample (see text). A clear transition from counterclockwise to clockwise circulation is observed, consistent with the sign change of the bulk topological invariant.}
\end{figure*}

To probe these two phases, we study the surface states associated with the bulk topological invariants \cite{hasan2010topological}. Accordingly, a finite sample with counterclockwise waveguide chirality taken from the $W=+1$ ($W=-1$) phases should possess a single counterclockwise (clockwise) edge mode traversing the gap centered on $\beta=0$. Figure \ref{fig_bandstructure} shows the tight binding band structures computed for each of the two phases using a strip geometry that is periodic in one direction and finite in the other. We see that in both cases the system is gapped with a single edge mode traversing the gap. The edge mode group velocity is reversed between the two cases, in agreement with the opposite sign of the associated bulk invariants.

To observe these edge modes experimentally, we fabricate two honeycomb lattices that each form a triangle with 17 waveguides on a side.  Each side is terminated at a zig-zag edge. The structures are written in borosilicate glass with $n_0 = 1.473$ and $\delta n = 2.8\times10^{-3}$ using the femtosecond direct write technique \cite{davis1995writing}. We set the lattice constant to $a=22\sqrt{3}\mu m$. The waveguides have diameters of $7\mu m$ and $10.7\mu m$ along the $x$- and $y$-directions and the sample length is $14cm$. The helix parameters for the two samples -- which we will refer to as samples I and II -- are respectively given by $(R,Z)=(20\mu m, 1.0cm)$ and $(R,Z) = (106\mu m, 2.4cm)$. In both samples, the helices are fabricated with counterclockwise helix chirality. For the purposes of comparing sample I with the sample in reference \cite{rechtsman2013photonic}, please note that reference \cite{rechtsman2013photonic} uses clockwise helix chirality. The locations of these samples on the phase diagram are wavelength dependent and are shown in Figure \ref{fig_phasediag} for wavelengths in the range $1480$-$1600$nm. Note that for the large-radius sample, we have compensated for the additional loss that would be introduced upon increasing the helix radius by also increasing the period. As a result, the system lies well below the high-loss region, but as a by-product of attempting to satisfy the conflicting goals of moving to a new topological phase while minimizing losses, we have arrived in a counterintuitive regime where the diameter of a single helix is larger than the entire lattice.

To excite the edge modes, we shift the waveguides at the corners of the triangle so that their nearest neighbor spacing is a factor of $1.25$ larger than the nearest neighbor spacing defining the lattice. In the weak coupling limit, these waveguides couple primarily to modes centered around $\beta=0$ and hence excite the edge modes that cross the gap. We thus inject light at the corners, allow it to propagate through the structure, and then image it at the output facet. The results are shown in Figure \ref{fig_output}, where we see a clear transition from counterclockwise to clockwise propagation. 

While the propagation distance of the light is fixed for a given sample, we can effectively image the light at different stages of propagation by varying the group velocity of the edge mode. To accomplish this, we note that the tight-binding coupling constant is an increasing function of wavelength and, as a result, variation of the group velocity can be implemented via a wavelength sweep. Such a wavelength sweep simultaneously changes the coupling and shifts the system in the topological phase diagram along the paths shown in Figure \ref{fig_phasediag}. In our experiment, we implement a sweep from $1480$-$1600$nm. Over this range, the system remains in the same topological phase while exhibiting an increased group velocity for larger wavelengths. As a result, the transverse distance traveled by the edge modes arriving at the output facet is observed to increase with wavelength (see Figure \ref{fig_output} and Supplemental Material \cite{suppmat}).

In summary, we have considered a honeycomb array of helical waveguides operating in the paraxial limit and experimentally shown that we can use the amplitude degree of freedom to tune the waveguide array between topological phases with opposite winding number. We observed these phases by direct imaging of the associated chiral edge modes in a system with finite geometry. We avoided the non-trivial problem of bending loss encountered in previous experimental studies by effectively stretching the helices along the $z$-direction so as to lower the helix frequency while drastically increasing the helix radius in a way that simultaneously increases the waveguide curvature radius while keeping the system in the new topological phase.  This result goes beyond the photonic context discussed here in the sense that it may be applied to Floquet phases of two-dimensional solid-state materials (e.g., graphene).  In particular, these ideas have direct application to the mitigation of heating and the engineering of topological phases in the intermediate and strong-driving regimes of Floquet systems. We close by mentioning a related concurrent work \cite{concurrentIrvine} in which a topological transition was experimentally realized in a mechanical system consisting of a collection of coupled gyroscopes.

\begin{acknowledgments}
M.C.R. acknowledges the National Science Foundation under award number ECCS-1509546, the Charles E. Kaufman Foundation, a supporting organization of the Pittsburgh Foundation, and the Alfred P. Sloan Foundation under fellowship number FG-2016-6418. K.P.C. acknowledges the National Science Foundation under award numbers ECCS-1509199 and DMS-1620218. 
\end{acknowledgments}

\bibliography{transitionbib}

\begin{thebibliography}{32}%
\makeatletter
\providecommand \@ifxundefined [1]{%
 \@ifx{#1\undefined}
}%
\providecommand \@ifnum [1]{%
 \ifnum #1\expandafter \@firstoftwo
 \else \expandafter \@secondoftwo
 \fi
}%
\providecommand \@ifx [1]{%
 \ifx #1\expandafter \@firstoftwo
 \else \expandafter \@secondoftwo
 \fi
}%
\providecommand \natexlab [1]{#1}%
\providecommand \enquote  [1]{``#1''}%
\providecommand \bibnamefont  [1]{#1}%
\providecommand \bibfnamefont [1]{#1}%
\providecommand \citenamefont [1]{#1}%
\providecommand \href@noop [0]{\@secondoftwo}%
\providecommand \href [0]{\begingroup \@sanitize@url \@href}%
\providecommand \@href[1]{\@@startlink{#1}\@@href}%
\providecommand \@@href[1]{\endgroup#1\@@endlink}%
\providecommand \@sanitize@url [0]{\catcode `\\12\catcode `\$12\catcode
  `\&12\catcode `\#12\catcode `\^12\catcode `\_12\catcode `\%12\relax}%
\providecommand \@@startlink[1]{}%
\providecommand \@@endlink[0]{}%
\providecommand \url  [0]{\begingroup\@sanitize@url \@url }%
\providecommand \@url [1]{\endgroup\@href {#1}{\urlprefix }}%
\providecommand \urlprefix  [0]{URL }%
\providecommand \Eprint [0]{\href }%
\providecommand \doibase [0]{http://dx.doi.org/}%
\providecommand \selectlanguage [0]{\@gobble}%
\providecommand \bibinfo  [0]{\@secondoftwo}%
\providecommand \bibfield  [0]{\@secondoftwo}%
\providecommand \translation [1]{[#1]}%
\providecommand \BibitemOpen [0]{}%
\providecommand \bibitemStop [0]{}%
\providecommand \bibitemNoStop [0]{.\EOS\space}%
\providecommand \EOS [0]{\spacefactor3000\relax}%
\providecommand \BibitemShut  [1]{\csname bibitem#1\endcsname}%
\let\auto@bib@innerbib\@empty
%</preamble>
\bibitem [{\citenamefont {Haldane}\ and\ \citenamefont
  {Raghu}(2008)}]{haldane2008possible}%
  \BibitemOpen
  \bibfield  {author} {\bibinfo {author} {\bibfnamefont {F.~D.~M.}\
  \bibnamefont {Haldane}}\ and\ \bibinfo {author} {\bibfnamefont
  {S.}~\bibnamefont {Raghu}},\ }\href {\doibase 10.1103/PhysRevLett.100.013904}
  {\bibfield  {journal} {\bibinfo  {journal} {Phys. Rev. Lett.}\ }\textbf
  {\bibinfo {volume} {100}},\ \bibinfo {pages} {013904} (\bibinfo {year}
  {2008})}\BibitemShut {NoStop}%
\bibitem [{\citenamefont {Wang}\ \emph {et~al.}(2009)\citenamefont {Wang},
  \citenamefont {Chong}, \citenamefont {Joannopoulos},\ and\ \citenamefont
  {Soljacic}}]{wang2009observation}%
  \BibitemOpen
  \bibfield  {author} {\bibinfo {author} {\bibfnamefont {Z.}~\bibnamefont
  {Wang}}, \bibinfo {author} {\bibfnamefont {Y.}~\bibnamefont {Chong}},
  \bibinfo {author} {\bibfnamefont {J.~D.}\ \bibnamefont {Joannopoulos}}, \
  and\ \bibinfo {author} {\bibfnamefont {M.}~\bibnamefont {Soljacic}},\
  }\href@noop {} {\bibfield  {journal} {\bibinfo  {journal} {Nature}\ }\textbf
  {\bibinfo {volume} {461}},\ \bibinfo {pages} {772} (\bibinfo {year}
  {2009})}\BibitemShut {NoStop}%
\bibitem [{\citenamefont {Rechtsman}\ \emph {et~al.}(2013)\citenamefont
  {Rechtsman}, \citenamefont {Zeuner}, \citenamefont {Plotnik}, \citenamefont
  {Lumer}, \citenamefont {Podolsky}, \citenamefont {Dreisow}, \citenamefont
  {Nolte}, \citenamefont {Segev},\ and\ \citenamefont
  {Szameit}}]{rechtsman2013photonic}%
  \BibitemOpen
  \bibfield  {author} {\bibinfo {author} {\bibfnamefont {M.~C.}\ \bibnamefont
  {Rechtsman}}, \bibinfo {author} {\bibfnamefont {J.~M.}\ \bibnamefont
  {Zeuner}}, \bibinfo {author} {\bibfnamefont {Y.}~\bibnamefont {Plotnik}},
  \bibinfo {author} {\bibfnamefont {Y.}~\bibnamefont {Lumer}}, \bibinfo
  {author} {\bibfnamefont {D.}~\bibnamefont {Podolsky}}, \bibinfo {author}
  {\bibfnamefont {F.}~\bibnamefont {Dreisow}}, \bibinfo {author} {\bibfnamefont
  {S.}~\bibnamefont {Nolte}}, \bibinfo {author} {\bibfnamefont
  {M.}~\bibnamefont {Segev}}, \ and\ \bibinfo {author} {\bibfnamefont
  {A.}~\bibnamefont {Szameit}},\ }\href {http://dx.doi.org/10.1038/nature12066}
  {\bibfield  {journal} {\bibinfo  {journal} {Nature}\ }\textbf {\bibinfo
  {volume} {496}},\ \bibinfo {pages} {196} (\bibinfo {year}
  {2013})}\BibitemShut {NoStop}%
\bibitem [{\citenamefont {Hafezi}\ \emph {et~al.}(2013)\citenamefont {Hafezi},
  \citenamefont {Mittal}, \citenamefont {Fan}, \citenamefont {Migdall},\ and\
  \citenamefont {Taylor}}]{hafezi2013imaging}%
  \BibitemOpen
  \bibfield  {author} {\bibinfo {author} {\bibfnamefont {M.}~\bibnamefont
  {Hafezi}}, \bibinfo {author} {\bibfnamefont {S.}~\bibnamefont {Mittal}},
  \bibinfo {author} {\bibfnamefont {J.}~\bibnamefont {Fan}}, \bibinfo {author}
  {\bibfnamefont {A.}~\bibnamefont {Migdall}}, \ and\ \bibinfo {author}
  {\bibfnamefont {J.~M.}\ \bibnamefont {Taylor}},\ }\href@noop {} {\bibfield
  {journal} {\bibinfo  {journal} {Nature Photonics}\ }\textbf {\bibinfo
  {volume} {7}},\ \bibinfo {pages} {1001} (\bibinfo {year} {2013})}\BibitemShut
  {NoStop}%
\bibitem [{\citenamefont {Lu}\ \emph {et~al.}(2014)\citenamefont {Lu},
  \citenamefont {Joannopoulos},\ and\ \citenamefont
  {Soljacic}}]{lu2014topological}%
  \BibitemOpen
  \bibfield  {author} {\bibinfo {author} {\bibfnamefont {L.}~\bibnamefont
  {Lu}}, \bibinfo {author} {\bibfnamefont {J.~D.}\ \bibnamefont
  {Joannopoulos}}, \ and\ \bibinfo {author} {\bibfnamefont {M.}~\bibnamefont
  {Soljacic}},\ }\href@noop {} {\bibfield  {journal} {\bibinfo  {journal} {Nat
  Photon}\ }\textbf {\bibinfo {volume} {8}},\ \bibinfo {pages} {821} (\bibinfo
  {year} {2014})}\BibitemShut {NoStop}%
\bibitem [{\citenamefont {Cheng}\ \emph {et~al.}(2016)\citenamefont {Cheng},
  \citenamefont {Jouvaud}, \citenamefont {Ni}, \citenamefont {Mousavi},
  \citenamefont {Genack},\ and\ \citenamefont {Khanikaev}}]{cheng2016robust}%
  \BibitemOpen
  \bibfield  {author} {\bibinfo {author} {\bibfnamefont {X.}~\bibnamefont
  {Cheng}}, \bibinfo {author} {\bibfnamefont {C.}~\bibnamefont {Jouvaud}},
  \bibinfo {author} {\bibfnamefont {X.}~\bibnamefont {Ni}}, \bibinfo {author}
  {\bibfnamefont {S.~H.}\ \bibnamefont {Mousavi}}, \bibinfo {author}
  {\bibfnamefont {A.~Z.}\ \bibnamefont {Genack}}, \ and\ \bibinfo {author}
  {\bibfnamefont {A.~B.}\ \bibnamefont {Khanikaev}},\ }\href
  {http://dx.doi.org/10.1038/nmat4573} {\ \textbf {\bibinfo {volume} {15}},\
  \bibinfo {pages} {542} (\bibinfo {year} {2016})}\BibitemShut {NoStop}%
\bibitem [{\citenamefont {Jotzu}\ \emph {et~al.}(2014)\citenamefont {Jotzu},
  \citenamefont {Messer}, \citenamefont {Desbuquois}, \citenamefont {Lebrat},
  \citenamefont {Uehlinger}, \citenamefont {Greif},\ and\ \citenamefont
  {Esslinger}}]{jotzu2014experimental}%
  \BibitemOpen
  \bibfield  {author} {\bibinfo {author} {\bibfnamefont {G.}~\bibnamefont
  {Jotzu}}, \bibinfo {author} {\bibfnamefont {M.}~\bibnamefont {Messer}},
  \bibinfo {author} {\bibfnamefont {R.}~\bibnamefont {Desbuquois}}, \bibinfo
  {author} {\bibfnamefont {M.}~\bibnamefont {Lebrat}}, \bibinfo {author}
  {\bibfnamefont {T.}~\bibnamefont {Uehlinger}}, \bibinfo {author}
  {\bibfnamefont {D.}~\bibnamefont {Greif}}, \ and\ \bibinfo {author}
  {\bibfnamefont {T.}~\bibnamefont {Esslinger}},\ }\href@noop {} {\bibfield
  {journal} {\bibinfo  {journal} {Nature}\ }\textbf {\bibinfo {volume} {515}},\
  \bibinfo {pages} {237} (\bibinfo {year} {2014})}\BibitemShut {NoStop}%
\bibitem [{\citenamefont {Aidelsburger}\ \emph {et~al.}(2015)\citenamefont
  {Aidelsburger}, \citenamefont {Lohse}, \citenamefont {Schweizer},
  \citenamefont {Atala}, \citenamefont {Barreiro}, \citenamefont {Nascimbene},
  \citenamefont {Cooper}, \citenamefont {Bloch},\ and\ \citenamefont
  {Goldman}}]{aidelsburger2015measuring}%
  \BibitemOpen
  \bibfield  {author} {\bibinfo {author} {\bibfnamefont {M.}~\bibnamefont
  {Aidelsburger}}, \bibinfo {author} {\bibfnamefont {M.}~\bibnamefont {Lohse}},
  \bibinfo {author} {\bibfnamefont {C.}~\bibnamefont {Schweizer}}, \bibinfo
  {author} {\bibfnamefont {M.}~\bibnamefont {Atala}}, \bibinfo {author}
  {\bibfnamefont {J.~T.}\ \bibnamefont {Barreiro}}, \bibinfo {author}
  {\bibfnamefont {S.}~\bibnamefont {Nascimbene}}, \bibinfo {author}
  {\bibfnamefont {N.~R.}\ \bibnamefont {Cooper}}, \bibinfo {author}
  {\bibfnamefont {I.}~\bibnamefont {Bloch}}, \ and\ \bibinfo {author}
  {\bibfnamefont {N.}~\bibnamefont {Goldman}},\ }\href@noop {} {\bibfield
  {journal} {\bibinfo  {journal} {Nat Phys}\ }\textbf {\bibinfo {volume}
  {11}},\ \bibinfo {pages} {162} (\bibinfo {year} {2015})}\BibitemShut
  {NoStop}%
\bibitem [{\citenamefont {Goldman}\ \emph {et~al.}(2016)\citenamefont
  {Goldman}, \citenamefont {Budich},\ and\ \citenamefont
  {Zoller}}]{goldman2016topological}%
  \BibitemOpen
  \bibfield  {author} {\bibinfo {author} {\bibfnamefont {N.}~\bibnamefont
  {Goldman}}, \bibinfo {author} {\bibfnamefont {J.~C.}\ \bibnamefont {Budich}},
  \ and\ \bibinfo {author} {\bibfnamefont {P.}~\bibnamefont {Zoller}},\
  }\href@noop {} {\bibfield  {journal} {\bibinfo  {journal} {Nat Phys}\
  }\textbf {\bibinfo {volume} {12}},\ \bibinfo {pages} {639} (\bibinfo {year}
  {2016})}\BibitemShut {NoStop}%
\bibitem [{\citenamefont {Prodan}\ and\ \citenamefont
  {Prodan}(2009)}]{prodan2009topological}%
  \BibitemOpen
  \bibfield  {author} {\bibinfo {author} {\bibfnamefont {E.}~\bibnamefont
  {Prodan}}\ and\ \bibinfo {author} {\bibfnamefont {C.}~\bibnamefont
  {Prodan}},\ }\href {\doibase 10.1103/PhysRevLett.103.248101} {\bibfield
  {journal} {\bibinfo  {journal} {Phys. Rev. Lett.}\ }\textbf {\bibinfo
  {volume} {103}},\ \bibinfo {pages} {248101} (\bibinfo {year}
  {2009})}\BibitemShut {NoStop}%
\bibitem [{\citenamefont {Zhang}\ \emph {et~al.}(2010)\citenamefont {Zhang},
  \citenamefont {Ren}, \citenamefont {Wang},\ and\ \citenamefont
  {Li}}]{zhang2010topological}%
  \BibitemOpen
  \bibfield  {author} {\bibinfo {author} {\bibfnamefont {L.}~\bibnamefont
  {Zhang}}, \bibinfo {author} {\bibfnamefont {J.}~\bibnamefont {Ren}}, \bibinfo
  {author} {\bibfnamefont {J.-S.}\ \bibnamefont {Wang}}, \ and\ \bibinfo
  {author} {\bibfnamefont {B.}~\bibnamefont {Li}},\ }\href@noop {} {\bibfield
  {journal} {\bibinfo  {journal} {Phys. Rev. Lett.}\ }\textbf {\bibinfo
  {volume} {105}},\ \bibinfo {pages} {225901} (\bibinfo {year}
  {2010})}\BibitemShut {NoStop}%
\bibitem [{\citenamefont {Kane}\ and\ \citenamefont
  {Lubensky}(2014)}]{kane2014topological}%
  \BibitemOpen
  \bibfield  {author} {\bibinfo {author} {\bibfnamefont {C.~L.}\ \bibnamefont
  {Kane}}\ and\ \bibinfo {author} {\bibfnamefont {T.~C.}\ \bibnamefont
  {Lubensky}},\ }\href@noop {} {\bibfield  {journal} {\bibinfo  {journal} {Nat
  Phys}\ }\textbf {\bibinfo {volume} {10}},\ \bibinfo {pages} {39} (\bibinfo
  {year} {2014})}\BibitemShut {NoStop}%
\bibitem [{\citenamefont {Nash}\ \emph {et~al.}(2015)\citenamefont {Nash},
  \citenamefont {Kleckner}, \citenamefont {Read}, \citenamefont {Vitelli},
  \citenamefont {Turner},\ and\ \citenamefont {Irvine}}]{nash2015topological}%
  \BibitemOpen
  \bibfield  {author} {\bibinfo {author} {\bibfnamefont {L.~M.}\ \bibnamefont
  {Nash}}, \bibinfo {author} {\bibfnamefont {D.}~\bibnamefont {Kleckner}},
  \bibinfo {author} {\bibfnamefont {A.}~\bibnamefont {Read}}, \bibinfo {author}
  {\bibfnamefont {V.}~\bibnamefont {Vitelli}}, \bibinfo {author} {\bibfnamefont
  {A.~M.}\ \bibnamefont {Turner}}, \ and\ \bibinfo {author} {\bibfnamefont
  {W.~T.~M.}\ \bibnamefont {Irvine}},\ }\href {\doibase
  10.1073/pnas.1507413112} {\bibfield  {journal} {\bibinfo  {journal}
  {Proceedings of the National Academy of Sciences}\ }\textbf {\bibinfo
  {volume} {112}},\ \bibinfo {pages} {14495} (\bibinfo {year}
  {2015})}\BibitemShut {NoStop}%
\bibitem [{\citenamefont {Huber}(2016)}]{huber2016topological}%
  \BibitemOpen
  \bibfield  {author} {\bibinfo {author} {\bibfnamefont {S.~D.}\ \bibnamefont
  {Huber}},\ }\href@noop {} {\bibfield  {journal} {\bibinfo  {journal} {Nat
  Phys}\ }\textbf {\bibinfo {volume} {12}},\ \bibinfo {pages} {621} (\bibinfo
  {year} {2016})}\BibitemShut {NoStop}%
\bibitem [{\citenamefont {Szameit}\ and\ \citenamefont
  {Nolte}(2010)}]{szameit2010discrete}%
  \BibitemOpen
  \bibfield  {author} {\bibinfo {author} {\bibfnamefont {A.}~\bibnamefont
  {Szameit}}\ and\ \bibinfo {author} {\bibfnamefont {S.}~\bibnamefont
  {Nolte}},\ }\href@noop {} {\bibfield  {journal} {\bibinfo  {journal} {Journal
  of Physics B: Atomic, Molecular and Optical Physics}\ }\textbf {\bibinfo
  {volume} {43}},\ \bibinfo {pages} {163001} (\bibinfo {year}
  {2010})}\BibitemShut {NoStop}%
\bibitem [{\citenamefont {Oka}\ and\ \citenamefont
  {Aoki}(2009)}]{oka2009photovolatic}%
  \BibitemOpen
  \bibfield  {author} {\bibinfo {author} {\bibfnamefont {T.}~\bibnamefont
  {Oka}}\ and\ \bibinfo {author} {\bibfnamefont {H.}~\bibnamefont {Aoki}},\
  }\href {\doibase 10.1103/PhysRevB.79.081406} {\bibfield  {journal} {\bibinfo
  {journal} {Phys. Rev. B}\ }\textbf {\bibinfo {volume} {79}},\ \bibinfo
  {pages} {081406} (\bibinfo {year} {2009})}\BibitemShut {NoStop}%
\bibitem [{\citenamefont {Kitagawa}\ \emph {et~al.}(2010)\citenamefont
  {Kitagawa}, \citenamefont {Berg}, \citenamefont {Rudner},\ and\ \citenamefont
  {Demler}}]{kitagawa2010topological}%
  \BibitemOpen
  \bibfield  {author} {\bibinfo {author} {\bibfnamefont {T.}~\bibnamefont
  {Kitagawa}}, \bibinfo {author} {\bibfnamefont {E.}~\bibnamefont {Berg}},
  \bibinfo {author} {\bibfnamefont {M.}~\bibnamefont {Rudner}}, \ and\ \bibinfo
  {author} {\bibfnamefont {E.}~\bibnamefont {Demler}},\ }\href {\doibase
  10.1103/PhysRevB.82.235114} {\bibfield  {journal} {\bibinfo  {journal} {Phys.
  Rev. B}\ }\textbf {\bibinfo {volume} {82}},\ \bibinfo {pages} {235114}
  (\bibinfo {year} {2010})}\BibitemShut {NoStop}%
\bibitem [{\citenamefont {Lindner}\ \emph {et~al.}(2011)\citenamefont
  {Lindner}, \citenamefont {Refael},\ and\ \citenamefont
  {Galitski}}]{lindner2011floquet}%
  \BibitemOpen
  \bibfield  {author} {\bibinfo {author} {\bibfnamefont {N.~H.}\ \bibnamefont
  {Lindner}}, \bibinfo {author} {\bibfnamefont {G.}~\bibnamefont {Refael}}, \
  and\ \bibinfo {author} {\bibfnamefont {V.}~\bibnamefont {Galitski}},\ }\href
  {\doibase 10.1038/nphys1926} {\bibfield  {journal} {\bibinfo  {journal} {Nat
  Phys}\ }\textbf {\bibinfo {volume} {7}},\ \bibinfo {pages} {490} (\bibinfo
  {year} {2011})}\BibitemShut {NoStop}%
\bibitem [{\citenamefont {Gu}\ \emph {et~al.}(2011)\citenamefont {Gu},
  \citenamefont {Fertig}, \citenamefont {Arovas},\ and\ \citenamefont
  {Auerbach}}]{gu2011floquet}%
  \BibitemOpen
  \bibfield  {author} {\bibinfo {author} {\bibfnamefont {Z.}~\bibnamefont
  {Gu}}, \bibinfo {author} {\bibfnamefont {H.~A.}\ \bibnamefont {Fertig}},
  \bibinfo {author} {\bibfnamefont {D.~P.}\ \bibnamefont {Arovas}}, \ and\
  \bibinfo {author} {\bibfnamefont {A.}~\bibnamefont {Auerbach}},\ }\href
  {\doibase 10.1103/PhysRevLett.107.216601} {\bibfield  {journal} {\bibinfo
  {journal} {Phys. Rev. Lett.}\ }\textbf {\bibinfo {volume} {107}},\ \bibinfo
  {pages} {216601} (\bibinfo {year} {2011})}\BibitemShut {NoStop}%
\bibitem [{\citenamefont {Mikami}\ \emph {et~al.}(2016)\citenamefont {Mikami},
  \citenamefont {Kitamura}, \citenamefont {Yasuda}, \citenamefont {Tsuji},
  \citenamefont {Oka},\ and\ \citenamefont {Aoki}}]{mikami2016brillouin}%
  \BibitemOpen
  \bibfield  {author} {\bibinfo {author} {\bibfnamefont {T.}~\bibnamefont
  {Mikami}}, \bibinfo {author} {\bibfnamefont {S.}~\bibnamefont {Kitamura}},
  \bibinfo {author} {\bibfnamefont {K.}~\bibnamefont {Yasuda}}, \bibinfo
  {author} {\bibfnamefont {N.}~\bibnamefont {Tsuji}}, \bibinfo {author}
  {\bibfnamefont {T.}~\bibnamefont {Oka}}, \ and\ \bibinfo {author}
  {\bibfnamefont {H.}~\bibnamefont {Aoki}},\ }\href {\doibase
  10.1103/PhysRevB.93.144307} {\bibfield  {journal} {\bibinfo  {journal} {Phys.
  Rev. B}\ }\textbf {\bibinfo {volume} {93}},\ \bibinfo {pages} {144307}
  (\bibinfo {year} {2016})}\BibitemShut {NoStop}%
\bibitem [{\citenamefont {Rudner}\ \emph {et~al.}(2013)\citenamefont {Rudner},
  \citenamefont {Lindner}, \citenamefont {Berg},\ and\ \citenamefont
  {Levin}}]{rudner2013anomalous}%
  \BibitemOpen
  \bibfield  {author} {\bibinfo {author} {\bibfnamefont {M.~S.}\ \bibnamefont
  {Rudner}}, \bibinfo {author} {\bibfnamefont {N.~H.}\ \bibnamefont {Lindner}},
  \bibinfo {author} {\bibfnamefont {E.}~\bibnamefont {Berg}}, \ and\ \bibinfo
  {author} {\bibfnamefont {M.}~\bibnamefont {Levin}},\ }\href {\doibase
  10.1103/PhysRevX.3.031005} {\bibfield  {journal} {\bibinfo  {journal} {Phys.
  Rev. X}\ }\textbf {\bibinfo {volume} {3}},\ \bibinfo {pages} {031005}
  (\bibinfo {year} {2013})}\BibitemShut {NoStop}%
\bibitem [{\citenamefont {Leykam}\ \emph {et~al.}(2016)\citenamefont {Leykam},
  \citenamefont {Rechtsman},\ and\ \citenamefont
  {Chong}}]{leykam2016anamolous}%
  \BibitemOpen
  \bibfield  {author} {\bibinfo {author} {\bibfnamefont {D.}~\bibnamefont
  {Leykam}}, \bibinfo {author} {\bibfnamefont {M.~C.}\ \bibnamefont
  {Rechtsman}}, \ and\ \bibinfo {author} {\bibfnamefont {Y.~D.}\ \bibnamefont
  {Chong}},\ }\href {\doibase 10.1103/PhysRevLett.117.013902} {\bibfield
  {journal} {\bibinfo  {journal} {Phys. Rev. Lett.}\ }\textbf {\bibinfo
  {volume} {117}},\ \bibinfo {pages} {013902} (\bibinfo {year}
  {2016})}\BibitemShut {NoStop}%
\bibitem [{\citenamefont {Noh}\ \emph {et~al.}(2017)\citenamefont {Noh},
  \citenamefont {Huang}, \citenamefont {Leykam}, \citenamefont {Chong},
  \citenamefont {Chen},\ and\ \citenamefont {Rechtsman}}]{noh2017experimental}%
  \BibitemOpen
  \bibfield  {author} {\bibinfo {author} {\bibfnamefont {J.}~\bibnamefont
  {Noh}}, \bibinfo {author} {\bibfnamefont {S.}~\bibnamefont {Huang}}, \bibinfo
  {author} {\bibfnamefont {D.}~\bibnamefont {Leykam}}, \bibinfo {author}
  {\bibfnamefont {Y.~D.}\ \bibnamefont {Chong}}, \bibinfo {author}
  {\bibfnamefont {K.~P.}\ \bibnamefont {Chen}}, \ and\ \bibinfo {author}
  {\bibfnamefont {M.~C.}\ \bibnamefont {Rechtsman}},\ }\href
  {http://dx.doi.org/10.1038/nphys4072} {\bibfield  {journal} {\bibinfo
  {journal} {Nat Phys}\ }\textbf {\bibinfo {volume} {13}},\ \bibinfo {pages}
  {611} (\bibinfo {year} {2017})}\BibitemShut {NoStop}%
\bibitem [{\citenamefont {Haldane}(1988)}]{haldane1988model}%
  \BibitemOpen
  \bibfield  {author} {\bibinfo {author} {\bibfnamefont {F.~D.~M.}\
  \bibnamefont {Haldane}},\ }\href {\doibase 10.1103/PhysRevLett.61.2015}
  {\bibfield  {journal} {\bibinfo  {journal} {Phys. Rev. Lett.}\ }\textbf
  {\bibinfo {volume} {61}},\ \bibinfo {pages} {2015} (\bibinfo {year}
  {1988})}\BibitemShut {NoStop}%
\bibitem [{\citenamefont {Chang}\ \emph {et~al.}(2013)\citenamefont {Chang},
  \citenamefont {Zhang}, \citenamefont {Feng}, \citenamefont {Shen},
  \citenamefont {Zhang}, \citenamefont {Guo}, \citenamefont {Li}, \citenamefont
  {Ou}, \citenamefont {Wei}, \citenamefont {Wang}, \citenamefont {Ji},
  \citenamefont {Feng}, \citenamefont {Ji}, \citenamefont {Chen}, \citenamefont
  {Jia}, \citenamefont {Dai}, \citenamefont {Fang}, \citenamefont {Zhang},
  \citenamefont {He}, \citenamefont {Wang}, \citenamefont {Lu}, \citenamefont
  {Ma},\ and\ \citenamefont {Xue}}]{chang167experimental}%
  \BibitemOpen
  \bibfield  {author} {\bibinfo {author} {\bibfnamefont {C.-Z.}\ \bibnamefont
  {Chang}}, \bibinfo {author} {\bibfnamefont {J.}~\bibnamefont {Zhang}},
  \bibinfo {author} {\bibfnamefont {X.}~\bibnamefont {Feng}}, \bibinfo {author}
  {\bibfnamefont {J.}~\bibnamefont {Shen}}, \bibinfo {author} {\bibfnamefont
  {Z.}~\bibnamefont {Zhang}}, \bibinfo {author} {\bibfnamefont
  {M.}~\bibnamefont {Guo}}, \bibinfo {author} {\bibfnamefont {K.}~\bibnamefont
  {Li}}, \bibinfo {author} {\bibfnamefont {Y.}~\bibnamefont {Ou}}, \bibinfo
  {author} {\bibfnamefont {P.}~\bibnamefont {Wei}}, \bibinfo {author}
  {\bibfnamefont {L.-L.}\ \bibnamefont {Wang}}, \bibinfo {author}
  {\bibfnamefont {Z.-Q.}\ \bibnamefont {Ji}}, \bibinfo {author} {\bibfnamefont
  {Y.}~\bibnamefont {Feng}}, \bibinfo {author} {\bibfnamefont {S.}~\bibnamefont
  {Ji}}, \bibinfo {author} {\bibfnamefont {X.}~\bibnamefont {Chen}}, \bibinfo
  {author} {\bibfnamefont {J.}~\bibnamefont {Jia}}, \bibinfo {author}
  {\bibfnamefont {X.}~\bibnamefont {Dai}}, \bibinfo {author} {\bibfnamefont
  {Z.}~\bibnamefont {Fang}}, \bibinfo {author} {\bibfnamefont {S.-C.}\
  \bibnamefont {Zhang}}, \bibinfo {author} {\bibfnamefont {K.}~\bibnamefont
  {He}}, \bibinfo {author} {\bibfnamefont {Y.}~\bibnamefont {Wang}}, \bibinfo
  {author} {\bibfnamefont {L.}~\bibnamefont {Lu}}, \bibinfo {author}
  {\bibfnamefont {X.-C.}\ \bibnamefont {Ma}}, \ and\ \bibinfo {author}
  {\bibfnamefont {Q.-K.}\ \bibnamefont {Xue}},\ }\href {\doibase
  10.1126/science.1234414} {\bibfield  {journal} {\bibinfo  {journal}
  {Science}\ }\textbf {\bibinfo {volume} {340}},\ \bibinfo {pages} {167}
  (\bibinfo {year} {2013})}\BibitemShut {NoStop}%
\bibitem [{\citenamefont {Fukui}\ \emph {et~al.}(2005)\citenamefont {Fukui},
  \citenamefont {Hatsugai},\ and\ \citenamefont {Suzuki}}]{fukui2005chern}%
  \BibitemOpen
  \bibfield  {author} {\bibinfo {author} {\bibfnamefont {T.}~\bibnamefont
  {Fukui}}, \bibinfo {author} {\bibfnamefont {Y.}~\bibnamefont {Hatsugai}}, \
  and\ \bibinfo {author} {\bibfnamefont {H.}~\bibnamefont {Suzuki}},\ }\href
  {\doibase 10.1143/JPSJ.74.1674} {\bibfield  {journal} {\bibinfo  {journal}
  {Journal of the Physical Society of Japan}\ }\textbf {\bibinfo {volume}
  {74}},\ \bibinfo {pages} {1674} (\bibinfo {year} {2005})}\BibitemShut
  {NoStop}%
\bibitem [{\citenamefont {Marcuse}(1976)}]{marcuse1976curvature}%
  \BibitemOpen
  \bibfield  {author} {\bibinfo {author} {\bibfnamefont {D.}~\bibnamefont
  {Marcuse}},\ }\href {\doibase 10.1364/JOSA.66.000216} {\bibfield  {journal}
  {\bibinfo  {journal} {J. Opt. Soc. Am.}\ }\textbf {\bibinfo {volume} {66}},\
  \bibinfo {pages} {216} (\bibinfo {year} {1976})}\BibitemShut {NoStop}%
\bibitem [{\citenamefont {Bukov}\ \emph {et~al.}(2015)\citenamefont {Bukov},
  \citenamefont {D'Alessio},\ and\ \citenamefont
  {Polkovnikov}}]{bukov2015universal}%
  \BibitemOpen
  \bibfield  {author} {\bibinfo {author} {\bibfnamefont {M.}~\bibnamefont
  {Bukov}}, \bibinfo {author} {\bibfnamefont {L.}~\bibnamefont {D'Alessio}}, \
  and\ \bibinfo {author} {\bibfnamefont {A.}~\bibnamefont {Polkovnikov}},\
  }\href {\doibase 10.1080/00018732.2015.1055918} {\bibfield  {journal}
  {\bibinfo  {journal} {Advances in Physics}\ }\textbf {\bibinfo {volume}
  {64}},\ \bibinfo {pages} {139} (\bibinfo {year} {2015})}\BibitemShut
  {NoStop}%
\bibitem [{\citenamefont {Hasan}\ and\ \citenamefont
  {Kane}(2010)}]{hasan2010topological}%
  \BibitemOpen
  \bibfield  {author} {\bibinfo {author} {\bibfnamefont {M.~Z.}\ \bibnamefont
  {Hasan}}\ and\ \bibinfo {author} {\bibfnamefont {C.~L.}\ \bibnamefont
  {Kane}},\ }\href {\doibase 10.1103/RevModPhys.82.3045} {\bibfield  {journal}
  {\bibinfo  {journal} {Rev. Mod. Phys.}\ }\textbf {\bibinfo {volume} {82}},\
  \bibinfo {pages} {3045} (\bibinfo {year} {2010})}\BibitemShut {NoStop}%
\bibitem [{\citenamefont {Davis}\ \emph {et~al.}(1996)\citenamefont {Davis},
  \citenamefont {Miura}, \citenamefont {Sugimoto},\ and\ \citenamefont
  {Hirao}}]{davis1995writing}%
  \BibitemOpen
  \bibfield  {author} {\bibinfo {author} {\bibfnamefont {K.~M.}\ \bibnamefont
  {Davis}}, \bibinfo {author} {\bibfnamefont {K.}~\bibnamefont {Miura}},
  \bibinfo {author} {\bibfnamefont {N.}~\bibnamefont {Sugimoto}}, \ and\
  \bibinfo {author} {\bibfnamefont {K.}~\bibnamefont {Hirao}},\ }\href@noop {}
  {\bibfield  {journal} {\bibinfo  {journal} {Opt. Lett.}\ }\textbf {\bibinfo
  {volume} {21}},\ \bibinfo {pages} {1729} (\bibinfo {year}
  {1996})}\BibitemShut {NoStop}%
\bibitem [{sup()}]{suppmat}%
  \BibitemOpen
  \href@noop {} {}\bibinfo {note} {See Supplemental Material at URL for an
  animation showing the full wavelength sweep.}\BibitemShut {Stop}%
\bibitem [{\citenamefont {Mitchell}\ \emph {et~al.}()\citenamefont {Mitchell},
  \citenamefont {Nash},\ and\ \citenamefont {Irvine}}]{concurrentIrvine}%
  \BibitemOpen
  \bibfield  {author} {\bibinfo {author} {\bibfnamefont {N.~P.}\ \bibnamefont
  {Mitchell}}, \bibinfo {author} {\bibfnamefont {L.~M.}\ \bibnamefont {Nash}},
  \ and\ \bibinfo {author} {\bibfnamefont {W.~T.~M.}\ \bibnamefont {Irvine}},\
  }\href@noop {} {\bibinfo  {journal} {(submitted)}\ }\BibitemShut {NoStop}%
\end{thebibliography}%

\end{document}